# On the inverse and conventional magnetocaloric effects in MnRhAs single crystals

M. Balli*[1,2], D. Fruchart[1] and R. Zach[3]

[1] Département MCMF, Institut Néel, CNRS, BP 166, 38042 Grenoble, Cedex 9, France.

[2] Department of Physics, University of Sherbrooke, 2500 Blvd de l'Université, J1K2R1, QC Canada.

[3] Institute of Physics, Cracow University of Technology, Podchorążych 1, 30-084 Cracow, Poland

**Abstract**. We report on the magnetic and magnetocaloric properties of a MnRhAs single crystal. The ternary arsenide exhibits a rather complex magnetic behaviour. A first order metamagnetic type transition from antiferromagnetic ($AF_I$) to ferromagnetic ("F") states takes place at $T_T$ ~ 158 K, and a second order transition from ferromagnetic ("F") to antiferromagnetic ($AF_{II}$) states occurs at $T_C$ ~196 K, the paramagnetic state occurring at $T > T_N = 238$ K. Magnetic entropy changes were calculated using, Maxwell relation and Clausius-Clapeyron equation. Both approaches are compared and discussed. The $AF_I$-"F" transition in MnRhAs gives rise to an interestingly high level of negative magnetocaloric effect. Under a field change 0-1 T, the maximum magnetic entropy variation is about 3 J/kg K. For sufficiently high enough magnetic fields, the magnetocaloric working temperature range below 158 K can be covered. The "F"-$AF_{II}$ transition is accompanied by a relatively modest MCE (-2.3 J/kg K for 5 T at $T_C$ = 196 K), but it improves the working temperature span as well as the magnetocaloric properties. A minimum estimated refrigerant capacity of about 900 J/kg can be provided by a MnRhAs single crystal compound.

## 1. Introduction

Based on the magnetocaloric effect (MCE), magnetic refrigeration technology has attracted during the last decades a great interest due to its potential environmental benefits and the possibility to develop high efficiency devices [1, 2]. Moreover, the gap to be bridged in going from MCE to a practical device working at room temperature was very demanding due to the few number materials having a large magneto-thermal effect taking place around 300 K. Since its experimental observation[3], the conventional magnetocaloric effect principle has been widely implemented in low temperature applications [4].

Magnetic cooling has been promoted to ambient temperature range when Brown [5] unveiled in 1976 a room temperature reciprocating machine using the benchmark gadolinium (Gd) as refrigerant. However, a limited availability (leading to a high cost), a rather easy corrosion as well as a limited working temperature range have largely prevented commercialization of Gd-based MCE systems. Aiming to find an alternative for $Gd_{1-x}R_x$ (R = rare earth) alloys [6], the so-called giant MCE compound $Gd_5Ge_2Si_2$ was proposed by Pecharsky and Gschneidner [7]. This discovery rapidly led to a revival interest for magnetocaloric materials science. So, giant MCEs obtained under magnetic fields accessible via permanent magnets [8] were successively pointed out in well-known compounds such as MnAs compounds [9,10], $Fe_2P$-type compounds [11,12] and $LaFe_{13-x}Si_x$ [13-15]. On the other hand, magnetic refrigerants such as $RCo_2$-based compounds (R = rare earth) [16] were also proposed for low temperature applications (natural gas liquefaction for example). Following, a parallel effort was devoted to design new types of efficient magnetic refrigeration systems [17]. So, preindustrial systems were built up and tested [18-21] using the more recently discovered materials [22,23].

Among the reported refrigerant materials, Mn-based compounds have attracted a great interest due to their specific- structural and magnetic peculiarities [9-12,24]. In fact, manganese intermetallics often reveal specific magnetic trends since the effective exchange couplings is usually very sensitive to rather tenuous changes in the chemical bonds or/and structure characteristics. This offers the possibility to observe peculiar magnetic transformations as well as original but metastable magnetic configurations. So many first order magnetic to metamagnetic transitions were observed to take place at various temperatures in several series of Mn based compounds [9,10,12,25-29]. Such peculiarities open the possibility to handle easily those transitions when applying rather modest magnetic fields, accompanied in some cases by a negative MCE as for example for $Mn_3GaC$ and CoMnSi(Ge) [25-26]. Among many Mn-based intermetallics, the transition metal pnictides of formula MM'X (X = P, As, Sb…) exhibit very interesting fundamental properties [25,30-33]. Of the best series of magnetocaloric MM'X compounds, the phospho-arsenide $MnFeP_{\sim0.5}As_{\sim0.5}$, with some possible deviation of composition between Mn and Fe on one side and between P and As on the other, even substitution of few amounts of extra elements, led optimizing the MCEs or (and) to adjust both the temperature of application and the temperature span range. At present, aiming to avoid the unfair arsene chemical element, efforts are dedicated to the stabilization of non-As containing MM'X materials with e.g. X = ( P, Si, Ge) [34-37]. Here we report results related to a MnRhAs single crystal, a compound of fundamental interest mainly, but exhibiting both negative and positive magnetocaloric effects. The combination of both effects should make benefit in increasing the working temperature range in dedicated magnetic cooling systems.

MnRhAs compound belongs to the hexagonal $Fe_2P$-type crystal structure (space group P-62m) that displays two different metal sites, being a tetrahedral Fe (3f) site with four P as nearest neighbors (NN) atoms and a pyramidal site Fe (3g) with five P atoms as NN. Mn atoms occupy the pyramidal site and Rh atoms occupy the tetrahedral site with the lattice parameters of a ≈ 6.492 Å and c ≈ 3.715 Å [38] as measured by powder XRD experiments at 300 K. Along the c axis, Mn and Rh form alternate stacking layers, thus leading to a two-dimensional magnetic arrangement. MnRhAs was widely studied in the 90's, attracting fundamental interests due to his exotic magnetic state induced by temperature, by magnetic fields, even under high pressure [38-45].

Magnetization and neutron diffraction experiments [46, 47] first reveal that the Mn magnetic moments order with different types of magnetic arrangements, but in all cases by stacking 2D-ferromagnetic (001) planes. At low temperature, the magnetic moments of Mn display both in-plane $\mu_{xy}$ and perpendicular to plane $\mu_z$ component leading to a 4c-$AF_I$ magnetic periodicity, with $\mu_z$ = 1.2 $\mu_B$ and $\mu_{xy}$ = 3.33 $\mu_B$ for a total Mn moment of ≈ 3.55 $\mu_B$ measured at 6 K. Upon increasing temperature, at $T_T$ = 158 K, MnRhAs exhibits a 1$^{st}$ order type metamagnetic transition from there above non-collinear $AF_I$ to a complex ferromagnetic ("F") state with an abrupt enhancement of the magnetization at the transition region. At 178 K, $\mu_z$ = 1.2 $\mu_B$ and $\mu_{xy}$ = 2.0 $\mu_B$, the total moment is of ≈ 2.33 $\mu_B$. and for $T_T$ < T < $T_C$ (~ 196 K) the $\mu_z$ components are ferromagnetically coupled in between adjacent layers contrarily to the $\mu_{xy}$ ones that are antiferromagnetically coupled thus forming a "F" = F+$AF_{II}$ complex arrangement. Moreover in this temperature range, a weak magnetic moment of ≈ 0.2 $\mu_B$ appears on Rh atoms, oppositely polarized to the $\mu_z$ ferromagnetic component of Mn. Then, in between $T_C$ up to $T_N$ ~ 238 K, both the $\mu_z$ Mn-component and the antiparallel component of Rh-component vanish definitively, leading to a pure $AF_{II}$ collinear system of $\mu_{xy}$ Mn-components (~ 2.1 $\mu_B$ at 200 K) with a 2c-periodicity. However, in a short temperature range below $T_T$ where a metamagnetic behavior is found, neutron diffraction experiments on powder and single crystal [47] samples have allowed pointing out a modulated magnetic antiphase boundary system (not represented in Fig. 1) which super-period evolves rapidly with temperature [38 46, 47]. All here above described magnetic configurations are schematized in Figure 1 with their temperature range of stability.

According to the magnetic arrangements revealed by neutron diffraction, high magnetic field measurements show that the intermediate "F" phase of (F+$AF_{II}$) type is stabilized ($T_T$ ↘ and $T_C$ ↗), on the other hand, high pressure a.c. susceptibility measurements reveal different types of shift with $T_T$ ↗ and $T_C$ ↗ up to 4 GPa but again $T_T$ ↘ and $T_C$ ↗ for higher applied pressures [45]. Besides, X-ray diffraction experiments realized on powder and single crystals versus temperature did not reveal any modification in the crystal structure symmetry, but noticeable and abrupt changes of the a and c cell parameters occurring at $T_T$ and less marked and smooth changes at $T_C$ and $T_N$ [38,46].

Magnetoresistivity measurements realized on the single crystal have definitively demonstrated that the exchange interactions possess a significant electronic type character, also related to the 2D type of magnetic structure arrangements [39, 43-44]. Also some similarities in terms of electronic structure can be established with the FeRh [48] archetype of magnetocaloric compounds exhibiting a AF-F 1st order magnetic transition followed by a 2nd order one, where a reverse to Fe polarization involving the Rh 4d electrons support a global 3d-ferromagnetic long range ordering, respectively for Fe in FeRh and for Mn in MnRhAs [49]. Here, with MnRhAs, the situation is made more complicated probably due to the lowered symmetry from cubic Pm3m with FeRh to hexagonal P-62m with MnRhAs, leading to anisotropy contributions (exchange or magnetocrystalline anisotropy). A forthcoming paper will be dedicated to more specific analyses of the electronic structure of selected compounds exhibiting such 1st order AF-F reverse MCE.

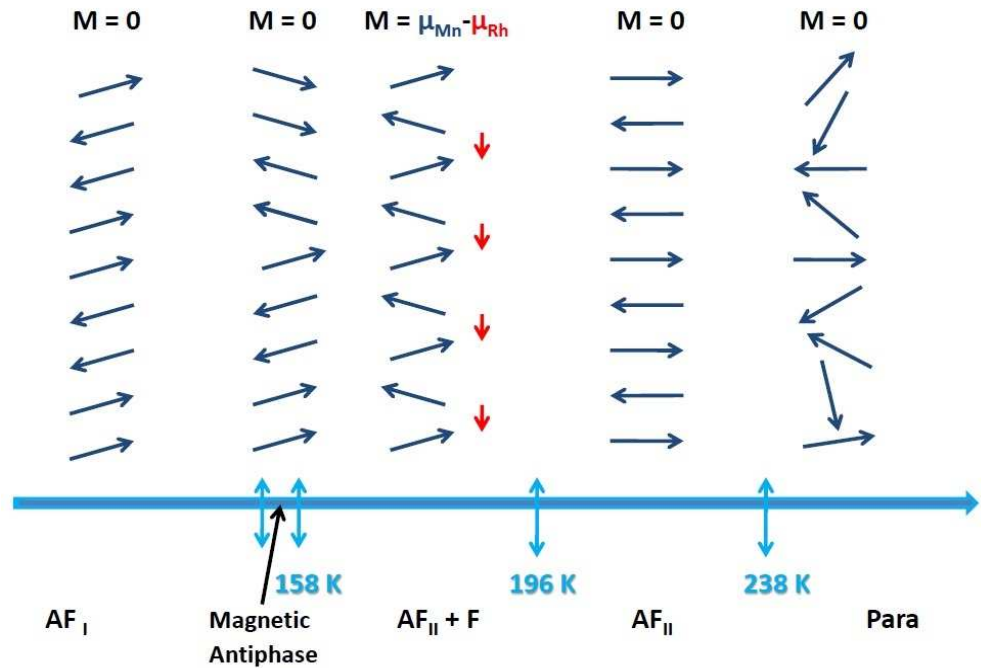

**Figure 1.** Schematic magnetic structures of MnRhAs at different temperatures. All (a, b) planes are ferromagnet layers. The $AF_I$ low temperature magnetic configuration extends on 4c (x,y) layers with both $\mu_{Mn}^{x,y}$ and $\mu_{Mn}^{z}$ components. The "F" = $AF_{II}$ + F configuration stabilized in between 158 and 196 K extends on 2 c (x,y) layers with both $\mu_{Mn}^{x,y}$ and $\mu_{Mn}^{z}$ components and an opposite $-\mu_{Rh}^{z}$ component. In a small temperature range of a few K, a magnetic antiphase system develops, integrating more and more $AF_{II}$ walls when temperature increases (H = 0), and at lower and lower temperature when the magnetic field H is increased. In between 196 K and the Néel temperature

of 238 K, the collinear antiferromagnetic configuration $AF_{II}$ is stabilized with a magnetic period extending on 2 c (x,y) layers.

## 2. Results and discussion

First a ~ 20 g ingot was prepared by solid state diffusion reaction of powdered elements (purity better than 4N) in sealed silica ampoules, followed by a first rapid melting up to ~ 1600 K. Then the here studied single crystal was prepared by using the Bridgman method applied to the ingot using a silica crucible in a high purity (5N5) argon atmosphere. XRD and neutron Laue patterns were realized to check for the single phase characteristic and to select the best part of the ingot in terms of single crystal state. Then a sphere of ~ 3 mm diameter was cut by spark erosion and by using a two axis rotating tube electrodes method (copper tube of 3.3 mm internal diameter) [39].

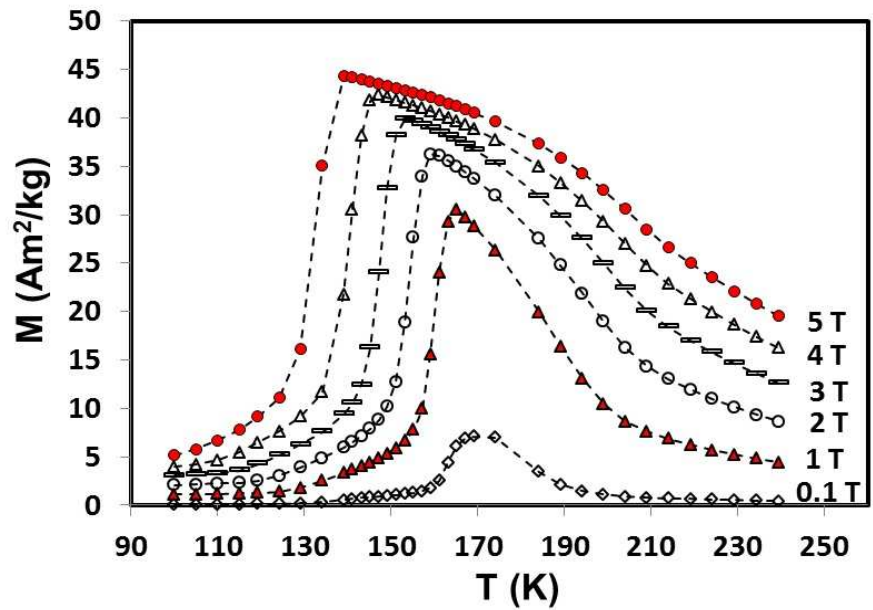

Figure 2. Temperature dependence of the MnRhAs single crystal magnetization, measured in different magnetic fields.

The measured magnetization curves along different axes of the MnRhAs single crystal were found almost similar showing that the magneto-crystalline anisotropy remains rather low [39, 46]. For this reason, the reported magnetization measurements were performed following an arbitrary orientation of the single crystal. The temperature dependence of the magnetization measured on the single crystalline sphere under several magnetic fields is shown in Figure 2. Increasing magnetic field, the $AF_I$-"F" transition region (where a modulated phase boundary was established taking place spontaneously) was found shifted to lower temperatures with a rate of about - 6.6 K/T and a linear relationship between the transition temperature $T_T$ and the magnetic field. The negative slope ($dT_T/dH$) and the increase of magnetization at $T_T$ are both indications for the existence of a negative or reverse magnetocaloric effect. However as shown in Figure 2, the first order character of the $AF_I$-"F" transition was retained even for high magnetic fields. In order to investigate the influence of the magnetic field on the $AF_I$-"F" phase transition, isothermal magnetization curves were recorded as represented between 110 and 163 K in Figure 3-a.

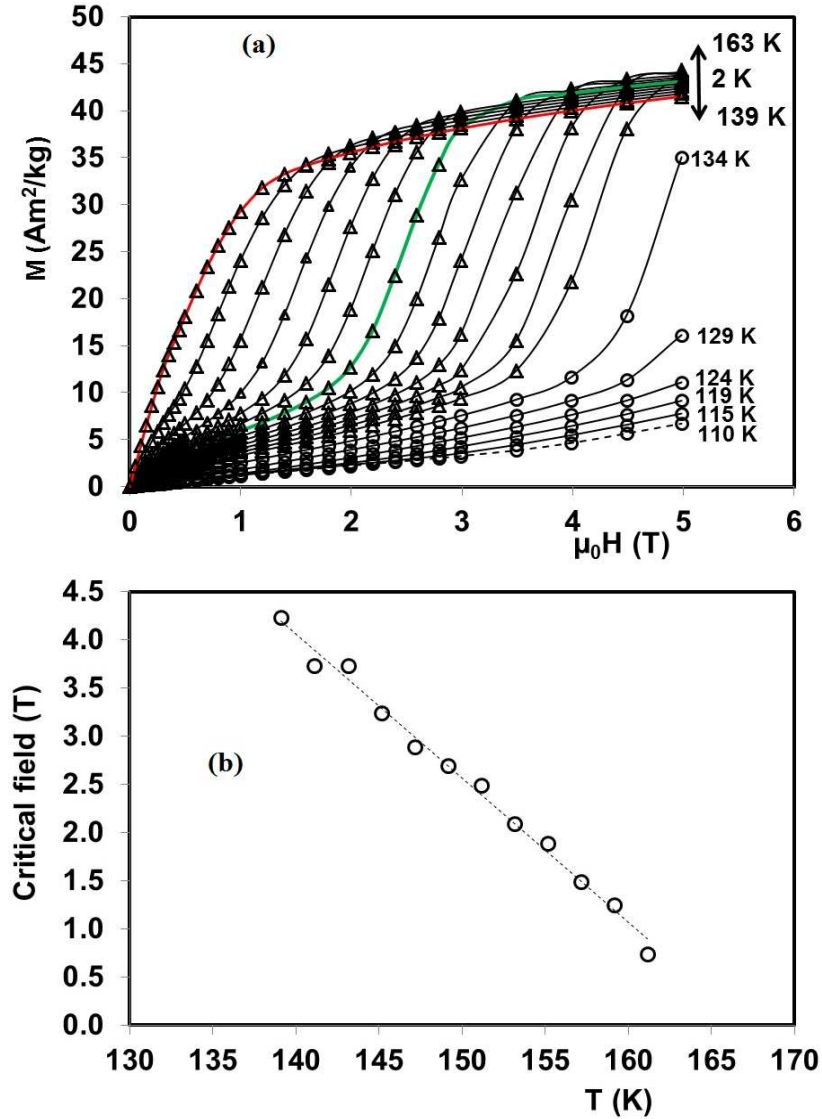

Figure 3. Metamagnetic characteristics of the MnRhAs single crystal. a - Isothermal magnetization curves in the AFI-"F" magnetic transition temperature range. b - Critical field HC variation measured versus temperature of a MnRhAs single crystal.

In the temperature range 161−139 K, the magnetization was found increasing almost linearly at low fields followed by an abrupt increase to a high magnetization state with the tendency to saturate, after overpassing a critical magnetic field $H_C$. This sudden change of magnetization reveals the possibility to reverse the magnetic ordering from $AF_I$ to "F" state by application of magnetic fields large enough suggesting the metamagnetic character of the transition. The critical field which is defined as the inflexion point of magnetization curve versus magnetic field within the transitional region is shown in Figure 3-b. In contrast to paramagnetic-ferromagnetic field-induced metamagnetic transformations [50], the $H_C$ value related to $AF_I$-"F" transition in MnRhAs increases almost linearly with a negative rate of about -0.15 T/K while temperature decreases. As shown in Figure 3-a, for temperatures lower than 139 K, magnetic fields higher than 5 T are needed to induce the metamagnetic transition. Isothermal M (H) magnetization measurements have been performed [38] under intense magnetic fields up to 15 T at 4.2, 101 and 130 K, respectively. According to the reported data, a critical field of about 10 T is needed to reverse magnetization ($AF_I$–"F") at 4.2 K. This suggests that, using a magnetic field of 10 T is large enough to induce metamagnetic transition between 160 K and 4.2 K. Consequently, a wide range of temperature (> 150 K) can be covered by using only a single material (MnRhAs) without the need for the combination of several materials with different phase transitions [51]. Considering Figures 2 and 4, a second transition from "F" to the high temperature $AF_{II}$ state is observed close to $T_C$ = 196 K. From Arrot's plots (not reported here) this transition is confirmed having a $2^{nd}$ order character. The measured transition temperatures $T_T$ = 158 K and $T_C$ = 196 K, remain in good agreement with those reported in Refs. 38 and 42. However, it is worth noting that the micro-magnetic structure of MnRhAs is more complex as determined by neutron diffraction [42, 43] and specific heat measurements [44] and detailed here above in the introduction.

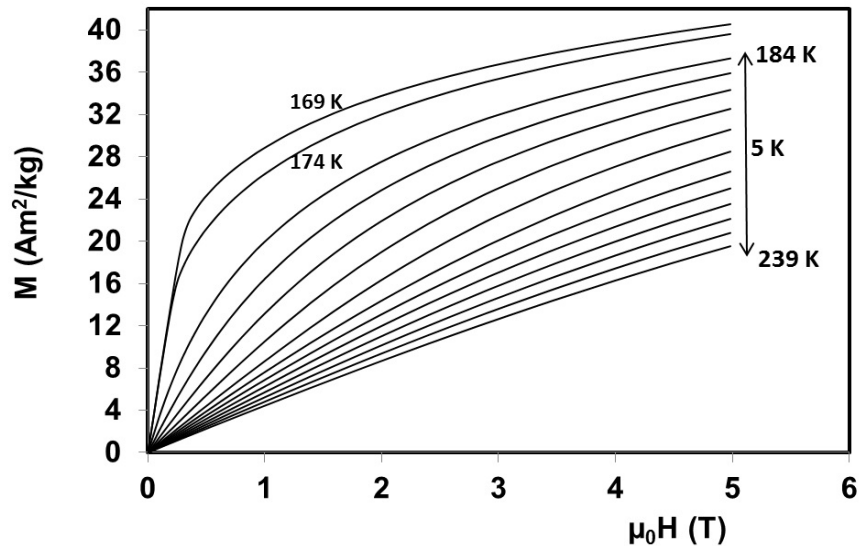

Figure 4. Isothermal magnetization curves of the MnRhAs single crystal above the AFI-"F" transition temperature TT.

The MCE represented by the isothermal entropy change and the adiabatic temperature change can be evaluated by measuring directly $\Delta T_{ad}$ or indirectly from magnetization and/or specific heat measurements. The Maxwell relation (MR) based approach is the common method used widely in the literature to determine $\Delta S$. This indirect method is considered as easy tool to evaluate fast the magnetocaloric performance of magnetic materials. However, the calculation of the MCE in materials with first order magnetic transition (FOMT) on the basis of MR was controverted leading to intense discussions in magnetocalorics community [50, 51, 53]. In materials with a large hysteresis resulting in phase separated state at the transition region, this method can overestimate largely the MCE. For low hysteresis, realistic values of the MCE could be obtained even in FOMT materials [50, 52-56]. We have considered first the MR to evaluate the MCE of MnRhAs at $T_T$. Furthermore, a series of isothermal magnetization curves was measured in the temperature ranges of interest, namely $AF_I$–"F" and "F"–$AF_{II}$ zones in external magnetic fields up 5 T (figures 3-a and 4). From the thermodynamic MR, the isothermal entropy change for external magnetic field varying from 0 to H is given as follow:

$$\Delta S(T, 0 \to H) = \int_0^H \left(\frac{\partial M}{\partial T}\right)_{H'} dH' \qquad (1)$$

Using magnetization measured at discrete values of magnetic field and temperature, one can calculate the entropy change through numerical integration of equation (1):

$$\Delta S = \sum_i \frac{M_{i+1} - M_i}{T_{i+1} - T_i} \Delta H_i \qquad (2)$$

where $M_{i+1}$ and $M_i$ are the magnetization values measured in a field $H$, at temperatures $T_{i+1}$ and $T_i$, respectively. As shown in equation (1), the nature of the MCE i.e. negative or conventional, is governed by the sign of $\left(\frac{\partial M}{\partial T}\right)$. The temperature dependence of the isothermal entropy change related to $AF_I$–"F" region is reported in Figure 5 for several magnetic fields. As it is shown, the sign of $\Delta S$ is positive giving rise to a negative MCE. This is because the application of an external magnetic field changes the magnetic state from an ordered phase ($AF_I$) to a less-ordered "F" = ($AF_{II}$+F) phase which increases the material entropy. In adiabatic conditions, the system compensates for this change by cooling down. When the magnetic field is suppressed, the magnetic moments return to their ordered state, entropy decreases and the material is forced to heat up. The MCE exhibited by MnRhAs compound remains in contrast with the conventional one reported in the most majority of materials (ferromagnets), but it is similar to that of $Mn_3GaC$ [25], $CoMnSi_{1-x}Ge_x$ [27] and FeRh [55] type of compounds. For a magnetic field variation of 0 to 1 T, 0 to 2 T and 0 to 5 T, the maximum changes of entropy are respectively about 3 J/kg K, 4 J/kg K and 5 J/kg K. For magnetic fields higher than 3 T, the ΔS vs T curves evidence anomalies (minimum) around 150 K. This is more probably attributed to the numerical integration process of the MR. Such behavior was observed in numerous materials exhibiting FOMT [25]. The maximum entropy change increases slightly with the magnetic field, while the peak width of ΔS-T curve increases at a rate of about 7.5 K/T.

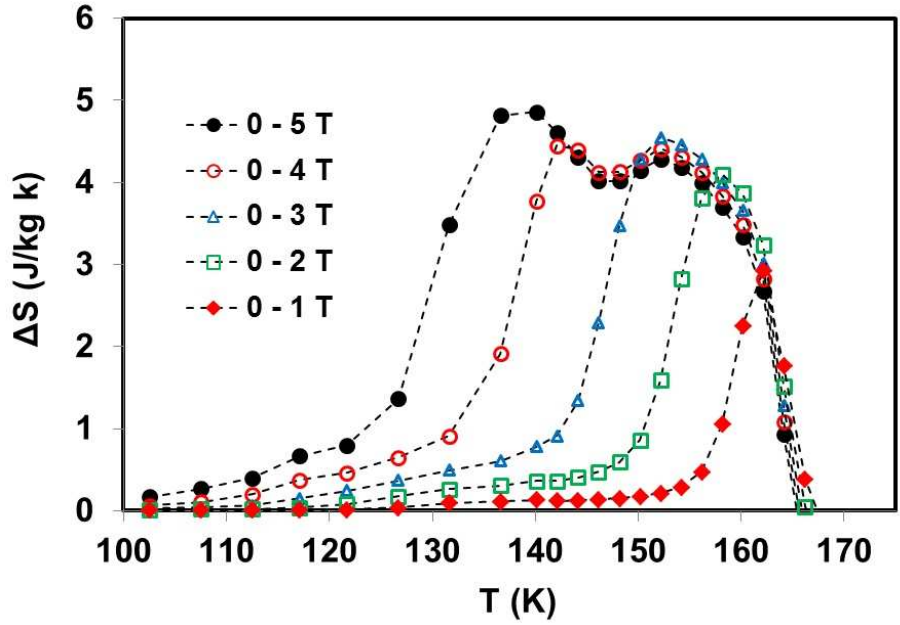

Figure 5. Isothermal entropy change resulting from the AFI-"F" transition of the MnRhAs single crystal as measured in several magnetic fields.

The entropy change can be also determined by Clausius-Clapeyron (C-C) equation, particularly when the high magnetization state has the tendency to saturate after the metamagnetic transformation [50, 54], which is the case of the present compound as shown in Figure 3-a. In C-C equation, ΔS is directly related to the metamagnetic transition through the magnetization jump at a given temperature. It is written as follow:

$$\Delta S = -\Delta M \frac{dH_C}{dT} = -\Delta M (\frac{dT_T}{dH})^{-1} \qquad (3)$$

ΔM can be determined as the difference of magnetization between the linear extrapolation of magnetization isotherms below and above the critical field $H_C$ which can be obtained from data of Figure 3-a. The value of $dH_C/dT$ is estimated from Figure 3-b to be about -0.15 T/K. The obtained ΔS versus temperature on MnRhAs compound by using the C-C equation is shown in Figure 6 where ΔS deduced from MR is also given for comparison. For MnRhAs a good agreement is observed between both approaches due essentially to the tendency of magnetization to saturate in high magnetization state. However, it is worth noting that the C-C equation allows calculate the change in the entropy only within the transition region. For a complete entropy change, C-C values should be completed by taking into account the region outside of phase transition[54].

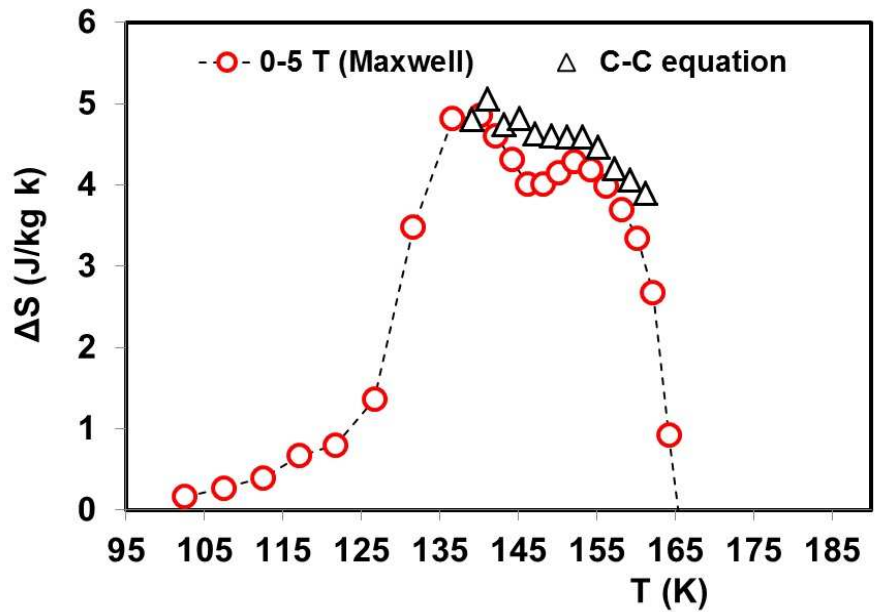

**Figure 6.** Isothermal entropy change of MnRhAs determined from the Clausius-Clapeyron equation (C-C). Maxwell relation (MR) data are also shown for comparison.

Looking at other previously investigated metamagnets with negative MCE, the entropy change found in MnRhAs is larger compared with that of $CoMnSi_{1-x}Ge_x$ where $\Delta S_{max}$ is about 1 J/kg K for 2 T [27]. Moreover, $Mn_3GaC$ [25] which presents similar AF-F transition temperature (160 K) has a $\Delta S_{max}$ (15 J/kg K for 2 T) more than three time larger if compared to MnRhAs (about 4 J/kg K for 2 T). The observed difference between these metamagnets can be understood on the basis of Eq. (3). The large entropy change observed in $Mn_3GaC$ is attributed to the large value of the magnetization jump which is about 70 $Am^2$/kg instead about 30 $Am^2$/kg for MnRhAs. In addition the high value of $dH_C/dT$ (-0.25 T/K) contributes also for the enhancement of the negative MCE in $Mn_3GaC$. Eventhough the $CoMnSi_{1-x}Ge_x$ system presents a high magnetization jump, the rate at which the metamagnetic transition temperature $dT_T/dH$ varies [25] with magnetic field is so far larger (-50 K/T) compared to -5 K/T for $Mn_3GaC$ and -6.6 K/T for MnRhAs. This reduces severly the entropy change in $CoMnSi_{1-x}Ge_x$[27]. In addition, the metamagnetic transition in $CoMnSi_{1-x}Ge_x$ takes places only for fields higher than 2 T, which is not the case of the MnRhAs and $Mn_3GaC$ compounds.

Up to now, the $Fe_{0.49}Rh_{0.51}$ [55] alloy can be considered as the most efficient magnetocaloric material. In addition to a conventional F-AF transition at high temperature, this alloy undergoes a field-induced AF-F FOMT at room temperature, which is accompanied by a large negative MCE. Under a magnetic field of 2 T, the temperature of a quenched sample $Fe_{0.49}Rh_{0.51}$ was decreased by about 13 K around $T_T = 313$ K. Unfortunately, the irreversibility of the giant MCE have excluded the $Fe_{0.49}Rh_{0.51}$ from implementation in magnetic cooling systems.

Concerning MnRhAs, the active magnetic refrigeration cycle used in the most majority of magnetic refrigerators, requires from the magnetocaloric refrigerant to exhibit interesting level of MCE on a large temperature span [56]. On the other hand, the Ericsson thermodynamic cycle implies a constant entropy change on the desired temperature range [6]. This demonstrates the diffuclty of using a single material in magnetic cooling technology. To deal with these requirements, a composite refrigerant built from several magnetocaloric materials was considered as a pertinent answer [51]. In this context, the temperature range between 4.2 (LHT) and 160 K can be covered by a MnRhAs type compound without the need to use composite refrigerants provided to use sufficiently high magnetic field. In fact, the possibility to retain the metamagnetic transtion ($AF_I$-"F") from $T_T = 158$ K up to helium temperature using a magnetic field of about 10 T, enables a broad range of working magnetocaloric temperatures by this compound. Noting that the $\Delta S$ is expected to vary slightly in 4.2 - 160 K temperature range. However, as shown in Figures 2 and 4, the compound MnRhAs exibits a second order transition $F+AF_{II}$ ("F") to $AF_{II}$ at 196 K which is accompanied by a moderate MCE at this temperature range. The measured entropy change at various magnetic fields in this zone is reported in Figure 7. Here we determined a smaller $\Delta S$ as a consequence of the nature of the magnetic transition and low magnetization. For a field change from 0 to 2 T and from 0 to 5 T, the corresponding maximum entropy changes are about -1.2 and -2.3 J/kg K. The existence of this second magnetocaloric zone, increases the working temperature span of MnRhAs and improves its refrigerant capacity (RC). This latter, is an important quantity used for the evaluation of materials magnetocaloric power. In addition to the value of the entropy change, RC takes into account the working temperature range of the selected refrigerant. It is defined as:

$$RC = \int_{FWHM} \Delta S(T) dT$$ where *FWHM* is the full width at half maximum. Taking into account both transitions ($AF_I$-

"F" and "F"-$AF_{II}$) and by giving "5 J/kg K" as minimum value of the entropy change in the $AF_I$-"F" region, a minimum estimated refrigerant capacity of about 900 J/kg can be extracted from the MnRhAs compound.

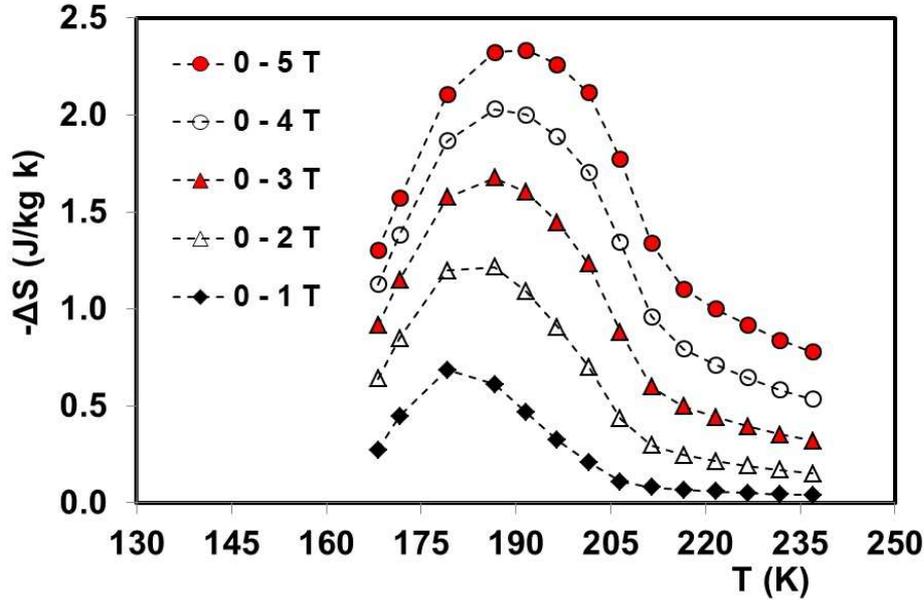

Figure 7. Isothermal entropy change corresponding to the "F" = (F+AFII) to the pure antiferromagnetic AFII configuration at a so-called Curie temperature $T_C$, and measured in different magnetic fields.

## 3. Conclusions

Magnetic and magnetocaloric properties of the single crystal MnRhAs have been studied. The complex nature of the magnetic structure gives rise to a negative and conventional magnetocaloric effects. Three magnetic phase transitions are observed in MnRhAs. A field induced metamagnetic transition from antiferromagnetic ($AF_I$) to a mixture of antiferromagnetic and ferromagnetic "F" = ($AF_{II}$+F) arrangements, takes place at $T_T$ = 158 K. A second magnetic phase transition from "F" = ($AF_{II}$+F) to pure antiferromagnetic $AF_{II}$ *s*tate occurs around $T_C$ = 196 K. Finally MnRhAs completely disorders in paramagnetic state at $T_N$ = 238 K.

Interesting levels of the negative MCE are observed below $T_T$ = 158 K on account of the metamagnetic character of the $AF_I$-"F" transition where modulated antiphase boundaries assist the overall magnetic transformation leading to excess entropy level. However, a rather good agreement was observed in the $AF_I$-"F" zone between MR and C-C equation, showing the ability of the latter for the evaluation of the entropy change in materials with such a metamagnetic transition. In contrast, the F+$AF_{II}$ to $AF_{II}$ transition leads to a relatively moderate MCE. Most of the entropy variation should be related to the depolarisation of the ferrimagnetic contributions and couplings between the $\mu_z$ components of Mn and that antiparallel found on Rh [38]. Here, in spite of the rather low value of the antiparallel coupled components (1.2 and -0.2 $\mu_B$ respectively on Mn and Rh), ΔS is expected having of pronounced electronic character related to the opposite 3d and 4d magnetic polarisation, as what was found for FeRh type of alloys as deduced from Ref. 48. However for a more precise analysis of the origin of the 4d magnetic contribution on Rh (probably leading here to an opposite Mn 3d response at $T_T$), precise band structure spin resolved calculation have to be developed for MnRhAs and related compounds. Also the effects of external pressure that leads to shift oppositely $T_T$ depending on the strength of the pressure [40], should be analysed in terms of the effective bonding forces taking place versus temperature in between the three elements Mn, Rh and As. Effectively between 130 and 300 K, precise crystal structure refinements demonstrate that the Rh-Rh and Mn-Mn distances remains unchanged (or weakly changed) contrary to the Mn-Rh ones that are affected by significant changes of some of the Mn-As and Rh-As bonding distances.

Anyway, combination of both the MCE effects found in MnRhAs is benefit in increasing both the RC and the working temperature range. Applying sufficient high magnetic fields around 10 T, a large magnetic cooling temperature span 4.2-160 K can be covered by a single compound MnRhAs - without the need to a composite material – expecting deliver a minimum refrigerant capacity of about ~ 900 J/kg. This is directly related to the effect of a strong enough magnetic field H to transform the magnetic low temperature $AF_I$ configuration to the higher temperature $F+AF_{II}$ configuration, thanks to the efficient H x $M_z$ coupling ($M_z = \mu_{Mn} - \mu_{Rh}$) that should be developed more and more within the long range antiphase boundary intermediate configuration.

Based on the here reported results, MnRhAs presents interesting magnetocaloric properties, particularly when using high enough magnetic fields. In addition, this work can also open the question on how improving again and tuning the magnetocaloric characteristics of MnRhAs, e.g. when either operating selected chemical substitutions (towards Mn(Rh,Ru)P i.e. monitoring the 4d-orbital contribution) or designing microstructured 2-D devices (i.e. layered deposits). On the other hand, for application tasks this study should be extended to a polycrystalline MnRhAs.